\documentclass[aps,prl,twocolumn,superscriptaddress,showpacs]{revtex4}
\usepackage{graphicx}
\usepackage{latexsym}
\usepackage{amssymb}
\usepackage{amsmath}
\usepackage{amsfonts}
\usepackage{bm}
\usepackage{multirow}
\usepackage{color}
\usepackage{comment}

\newcommand{\vect}[1]{{\bm{#1}}}

\DeclareMathAlphabet{\mathbbold}{U}{bbold}{m}{n}

\makeatletter
\newcommand\xleftrightarrow[2][]{%
\ext@arrow 9999{\longleftrightarrowfill@}{#1}{#2}}
\newcommand\longleftrightarrowfill@{%
\arrowfill@\leftarrow\relbar\rightarrow} \makeatother

\newcommand{\beq}{\begin{equation}}
\newcommand{\eeq}{\end{equation}}
\newcommand{\beqn}{\begin{eqnarray}}
\newcommand{\eeqn}{\end{eqnarray}}

\begin{document}

\title{Emergent Superconductivity in the weak Mott insulator phase of bilayer Graphene \\ Moir\'e
Superlattice}

\author{Xiao-Chuan Wu}

\author{Kelly Ann Pawlak}

\affiliation{Department of Physics, University of California,
Santa Barbara, CA 93106, USA}

\author{Chao-Ming Jian}
\affiliation{ Station Q, Microsoft Research, Santa Barbara,
California 93106-6105, USA} \affiliation{Kavli Institute of
Theoretical Physics, Santa Barbara, CA 93106, USA}

\author{Cenke Xu}
\affiliation{Department of Physics, University of California,
Santa Barbara, CA 93106, USA}

\date{\today}
\begin{abstract}

We propose a phenomenological understanding of the recently
discovered weak Mott insulator in the moir\'e superlattice of
twisted bilayer graphene, especially the emergent
superconductivity at low temperature within the weak Mott
insulator phase, namely while lowering temperature, the
longitudinal resistivity first grows below temperature $T_m$, but
then rapidly drops to zero at even lower temperature $T_c$. An
emergent superconductor in an insulator phase is very unusual.
Here we propose that this phenomenon is due to the pure
two-dimensional nature of the bilayer graphene moir\'e
superlattice. We also compare our results with other theories
proposed so far.

\end{abstract}

\maketitle

{\it --- Introduction}

Recent discovery of superconductivity (SC)~\cite{mag02} near a
weak Mott insulator (MI) phase in the graphene moir\'e
superlattice~\cite{mag01,chen2018gate} sheds new light on our
understanding of strongly correlated systems. This new system,
with unprecedented tunability, is an ideal experimental platform
to check our theoretical understandings. It is believed that the
nearly flat mini bands of the
system~\cite{flat1,flat2,flat3,flat4} play the major role in the
most interesting phenomena observed so far. Within the recent
theory works, Ref.~\onlinecite{xuleon,kivelson} described the
system with an effective two-orbital extended Hubbard model on a
triangular lattice near half-filling, the prediction of
Ref.~\onlinecite{xuleon} has been checked with numerical
methods~\cite{tripletcheck}; Ref.~\onlinecite{senthil,fu,lee}
described the system with a tight binding model on a honeycomb
lattice, while the electron Wannier functions strongly peak at the
triangular lattice sites. The main difference between these two
classes of models is that the latter models capture the physics
related to Dirac band crossings between a {\em pair} of flat
mini-bands. While at the doping where the SC and MI were observed,
$i.e.$ near half-filling within one of the mini-bands, it is not
clear that any symmetry protected Dirac point away from the band
plays a major role, unless one assumes a specific type of valley
order, which leads to extra Dirac crossings within the mini flat
band~\cite{senthil,lee}. But without compelling evidence of this
particular valley ordering in the MI phase, the qualitative
physics at the most relevant doping can potentially be captured by
the (simpler) effective triangular lattice models introduced in
Ref.~\onlinecite{xuleon,kivelson}. Especially since the activation
energy of the insulating phase is very low (4K)~\cite{mag01} even
compared with the narrow bandwidth and the effective Hubbard
interaction, this Mott insulator is rather weak and it is
conceivable that its insulating behavior can be understood based
solely on the electrons near the Fermi surface.

Nevertheless, the physics we discuss in the current work will be
largely independent of the details of the microscopic model. We
are going to focus on two peculiar and qualitative phenomena
observed in Ref.~\onlinecite{mag02}.

(1) The resistivity $R_{xx}(T)$ in Ref.~\onlinecite{mag02} shows
that at the Mott insulator doping, $R_{xx}(T)$ first increases
with lowering temperature below $T_m \sim 4-5$K (as one would
expect for an insulator), while rapidly drops to zero below
another temperature scale $T_c \sim 1$K. This feature means that
quite surprisingly the MI phase at very low energy scale still has
a superconductivity instability.

(2) Once the SC is suppressed by a weak external magnetic field,
the system becomes a normal MI with $R_{xx}(T)$ growing without
saturation at low temperature.

As we have mentioned the insulator phase in this system must be a
``weak" one, its activation gap is about the same as $k_B T_m$,
which is much lower than the estimated Hubbard interaction, even
with the large unit cell of the moir\'e structure. A weak Mott
insulator can be naturally understood based on physics around the
Fermi surface only. The electrons on the Fermi surface can be
gapped out by an order parameter at finite momentum through
folding of the Brillouin zone. When the amplitude of the order
parameter is weak, i.e. when the system is close to the
order-disorder quantum phase transition, only the ``hot spots" on
the Fermi surface connected by the momentum of the order parameter
are gapped out; but with a sufficiently strong order parameter and
its coupling to the electrons, the entire Fermi surface is gapped
out, and the system becomes an insulator, which can usually be
adiabatically connected to a strong Mott insulator at strong
coupling without any phase transition.

The simplest analogue of the physics described above is the
Hubbard model on the square lattice with nearest neighbor hopping
at exactly half filling. A weak Hubbard interaction will induce
the antiferromagnetic order at momentum $(\pi,\pi)$ and drive the
system into an insulator due to the Brillouin zone folding and
nesting of Fermi surface. And the insulator with weak Hubbard $U$
can be adiabatically connected to the insulator with large $U$,
where all the electrons are well localized on every site.

{\it --- Mechanism for weak MI and emergent SC}

Ref.~\onlinecite{xuleon,kivelson} both started with a two orbital
extended Hubbard model to understand the main experimental
observations of the moir\'e superlattice of twisted bilayer
graphene. The site of the triangular lattice is a patch of the
bilayer graphene with AA stacking. The two effective orbitals
correspond to the two valleys at the corners of the Brillouin zone
of the original honeycomb lattice. Both models in
Ref.~\onlinecite{xuleon,kivelson} have a SU(4) symmetry at the
leading order, and the SU(4) symmetry is broken by other
interactions such as the Hund's interaction.
Ref.~\onlinecite{xuleon,kivelson} chose a different sign for the
Hund's coupling, hence the former prefers a spin triplet and
valley singlet on every triangular lattice site, while the latter
prefers a spin singlet and valley order.

Here we first argue that the phenomena (1) and (2) mentioned above
can be both naturally explained within the framework of
Ref.~\onlinecite{xuleon}. A Hund's coupling chosen as
Ref.~\onlinecite{xuleon} will favor the two electrons on every
site in the Mott insulator phase to form a spin-1, with an
antiferromagnetic coupling between neighboring sites. The
frustrated nature of the triangular lattice will likely drive the
system into a spin density wave order. Even if we start with a
geometrically unfrustrated honeycomb lattice, the weakness of the
Mott insulator will also generate further neighbor spin
interaction and even multi-spin interactions which frustrate the
collinear magnetic order, and may as well lead to a spin density
wave (SDW). This SDW order connects different parts of the Fermi
surface through Brillouin zone folding. Phenomenon (2) suggests
that when a ``competing order" is suppressed and the SDW is
stabilized, the entire Fermi surface should be gapped out by the
SDW, i.e. there is no residual Fermi pocket left at the Fermi
surface, hence the amplitude of the SDW and its coupling to the
electrons are sufficiently strong. But let us not forget that the
system is purely two dimensional, hence with a full spin SU(2)
symmetry, the spins can never form a true long range order at
infinitesimal temperature. This situation is different from a
magnetic order close to its quantum critical point, in the sense
that close to a quantum critical point, both the amplitude and
direction of the magnetic order parameter will fluctuate strongly;
while in our case the amplitude of the SDW does not fluctuate
strongly, it is the direction of the order parameter that
modulates over a long correlation length scale $\xi(T)$.

\begin{figure}[tbp]
\begin{center}
\includegraphics[width=120pt]{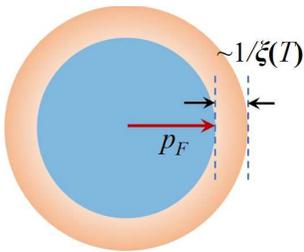}
\caption{The ``active" electrons within the thin momentum shell
around the Fermi surface with $|\vect{p} - \vect{p}_F| < \Lambda
\sim \hbar / \xi(T)$, which are insensitive to the background SDW
with finite correlation length $\xi(T)$, and hence can transport
electric charge and potentially form a SC. } \label{fig}
\end{center}
\end{figure}

A finite but long correlation length $\xi(T)$ implies that within
a thin momentum shell around the Fermi surface with $|\vect{p} -
\vect{k}_F| < \Lambda(T) \sim \hbar/\xi(T)$, the fermions will not
feel the background SDW order parameter with finite correlation
length. Rather than demonstrate this effect by detailed
calculations based on a microscopic model, one can visualize this
effect by simply coarse-graining the system, until $\hbar/\xi$
becomes the ultraviolet (UV) cut-off (thickness) of the momentum
shell around the Fermi surface following the standard
renormalization group picture of Fermi surface~\cite{ShankarRG},
and within this shell the electrons only see a very short range
correlated SDW, whose effects can be neglected. The electrons
within the thin shell are still ``active" and can transport
electric charge, or even form Cooper pairs (Fig.~\ref{fig}); while
the electrons outside this momentum shell will effectively view
the background SDW as a true long range order, and hence are
effectively ``gapped out". Based on the phenomenon (2), we know
that the gap induced by the SDW is strong enough when the SDW is
stabilized by an external field.

The active fermion density is proportional to the thickness of the
momentum shell $\Lambda(T) \sim \hbar/\xi(T)$. The correlation
length $\xi(T)$ of a SDW with a full SU(2) spin symmetry can be
estimated from the standard renormalization group calculation. Let
us take the noncollinear SDW as an example, which happens very
often in frustrated magnet (the experimental phenomena would also
be consistent with a collinear SDW at finite momentum). A
noncollinear SDW would break the entire SO(3) spin rotation group.
%A standard way of describing such SDW is to introduce the
%Schwinger boson $b_\alpha$ on every unit cell, such that
%$\vect{S}(\vect{r}) = \frac{1}{2} b^\dagger(\vect{r})
%\vect{\sigma} b(\vect{r})$, then to get a noncollinear SDW at
%finite momentum, $b(\vect{r})$ condenses at finite momentum
%$\vect{Q}$ and $-\vect{Q}$: \beqn b_\alpha(\vect{r}) \sim z_\alpha
%e^{i \vect{Q}\cdot \vect{r}} + w_\alpha e^{-i \vect{Q}\cdot
%\vect{r}}. \label{spinon} \eeqn A spin singlet pairing operator
%$\mathcal{P}\sim \epsilon_{\alpha\beta} z_\alpha w_\beta$ must
%condense to break the gauge degree of freedom of $b_\alpha$ from
%$U(1)$ down to $\mathbb{Z}_2$. The condensate of $\mathcal{P}$
%effectively makes $w_{\alpha} \sim \epsilon_{\alpha\beta}
%z^\ast_\beta$, thus the low energy physics of the condensate of
%$b_\alpha$ can be formulated in terms of $z_\alpha$ only.
The standard way of describing such SDW is to parameterize its
configuration manifold with two orthogonal vectors $\vect{n}_1$,
$\vect{n}_2$. It is convenient to introduce a SU(2) spinor field
$z = (z_1, z_2)^t$~\cite{senthilchubukov}, and \beqn \vect{n}_1
\sim \mathrm{Re}[z^t i\sigma^y \vect{\sigma} z], \ \ \ \vect{n}_2
\sim \mathrm{Im}[z^t i\sigma^y \vect{\sigma} z] \label{z}\eeqn $z
= (z_1, z_2)^t$ are complex bosonic fields at certain momentum
$\vect{Q}$, and subject to constraint $|z_1|^2 + |z_2|^2 = 1$. The
two component complex field $z_\alpha$ lives in a target manifold:
the three dimensional sphere $S^3$, and it must couple to a
$\mathbb{Z}_2$ gauge field~\cite{senthilchubukov}, and when
$z_\alpha$ condenses the ground state manifold is
$S^3/\mathbb{Z}_2$, which is identical to the ground state
manifold of a noncollinear SDW.

The finite temperature physics of the SDW is well described by the
nonlinear sigma model (NLSM) defined with the spinor $z_\alpha$
field: \beqn Z = \int D z_\alpha(x) \exp\left( - \int d^2x \
\frac{1}{2g} \sum_{\alpha}|\vect{\nabla} z_\alpha|^2 \right),
\eeqn where $g = k_B T/\rho_s$, and again $\rho_s$ is the spin
stiffness at zero temperature. The 2nd order renormalization group
(RG) equation of the coupling constant $g$ is \beqn \frac{dg}{d\ln
l} = \frac{1}{\pi}g^2 + O(g^3). \label{RG0} \eeqn For small $g$
(low temperature), the correlation length scales as: \beqn \xi(T)
\sim a_0 \exp\left( \frac{\pi \rho_s}{k_B T} \right),
\label{xi}\eeqn with an extra less important power-law function of
$T/\rho_s$ in the prefactor~\cite{nlsmref,sachdevbook}. $a_0$ is
the lattice constant of the moir\'e superlattice, $\rho_s$ is the
spin stiffness at zero temperature. This means that the energy
width of the momentum shell $v_f\Lambda(T)$ is much smaller than
the thermal energy $k_B T$ at low enough temperature $T$, hence
the electrons in this shell are fully thermally excited. Thus the
transport properties of these electrons can be captured by the
most classical theory of transport, such as the Drude theory. For
instance, the electric conductivity of the system is \beqn
\sigma(T) \sim \frac{n(T) e^2 \tau}{m^\ast}, \eeqn where $n(T)$ is
the density of electrons within this momentum shell, and it is
proportional to $\Lambda(T)$. Thus we can see that although there
is no true magnetic order at any finite temperature, due to the
rapidly decreasing density of active electrons within the momentum
shell, the resistivity $R_{xx}(T)$ will still rise with lowering
temperature, before the system becomes a SC.

%There is another physical picture of understanding these active
%electrons within the momentum shell. Assuming the magnetic order
%fluctuation has a much slower time scale compared with the
%electrons, then the electrons effectively move in a background of
%quenched (static) magnetic order domains, and roughly the size of
%each domain is $\xi(T)^2$. Within each domain the electrons are
%fully gapped, and the low energy electrons are localized at the
%domain walls. But since the domain has size $\xi(T)$, the momentum
%of the low energy electrons at the domain walls must be bounded by
%$\Lambda(T) \sim \hbar/\xi(T)$, which is the thickness of the
%aforementioned momentum shell.

At low temperature, the active electrons within the momentum shell
can still form a SC, which is consistent with the phenomenon (1)
mentioned above. But since the correlation length $\xi(T)$ grows
rapidly with lowering temperature, there are less and less active
electrons available for pairing, which is a sign of strong
competition between SC and the SDW. The SC transition temperature
$T_c$ for the active electrons can be estimated through the
standard BCS theory, under the assumption of a uniform gap
function around the Fermi surface (which is the case for almost
all the superconductors predicted in this system so far): \beqn
\frac{1}{J} = \int_0^{v_f \Lambda(T)} d\varepsilon
\frac{N}{\sqrt{\varepsilon^2 + \Delta^2}} \tanh \left(
\frac{\sqrt{\varepsilon^2 + \Delta^2}}{k_B T} \right), \label{SC}
\eeqn where $J$ represents the Heisenberg interaction on the
effective triangular lattice, which is the ``gluing force" for
superconductivity~\cite{xuleon}. In Eq.~\ref{SC} we have replaced
the UV cut-off of the standard BCS theory by $v_f \Lambda(T)$. As
always $N$ is the density of states around the Fermi surface,
which has been taken to be a constant. As we explained, at very
low temperature $v_f \Lambda(T)$ is much smaller than $k_B T$,
hence at $T_c$ ($\Delta = 0$), this equation can be simplified as
\beqn \frac{1}{NJ} = \frac{v_f \Lambda(T_c)}{k_B T_c}. \eeqn This
equation does not always have a solution, it only supports a
nonzero $T_c$ when $ NJ \gtrapprox\pi \rho_s a_0/(\hbar v_f)$.
Hence the system no longer has a BCS instability against
infinitesimal attractive interaction, the interaction $J$ needs to
be stronger than a critical strength.

{\it --- With weak anisotropy}

Once an external magnetic field is turned on (either inplane or
out-of-plane), the magnetic order will be more ``stabilized" at
low temperature because the spin symmetry is reduced to U(1),
which supports a quasi long range order with infinite correlation
length. In this case, the size of the momentum shell (and the
density of the active electrons) vanishes to zero, and there is no
room for SC.

The way a uniform Zeeman field couples to the spinor field
$z_\alpha$ depends on the symmetry of the noncollinear SDW, but it
will at least break the SO(3) symmetry down to U(1). A weak Zeeman
field $h$ will be renormalized to $h(l)$ at length scale $l$:
$l/a_0 \sim (h(l)/h)^{1/\delta} $, where $\delta$ is the scaling
dimension of $h$ in the NLSM; while at the same length scale the
coupling constant $g$ is renormalized according to Eq.~\ref{RG0}.
Comparison between the RG flow of $h(l)$ and $g(l)$ defines a
critical temperature $T'_c$: \beqn \left(
\frac{\rho_s}{h}\right)^{1/\delta} \sim \exp\left( \frac{\pi
\rho_s}{k_B T'_c} \right). \eeqn When $T \ll T_c'$, the coupling
constant $g(l)$ will still be small and perturbative when $h$
becomes nonperturbative compared with $\rho_s$, hence $g(l)$ stops
growing at a small value, and the system is in a quasi long range
ordered SDW phase; while when $T \gg T_c'$, the coupling constant
$g(l)$ becomes nonpeturbative before $h(l)$ could affect the RG
flow of Eq.~\ref{RG0}, and the system is in the disordered phase.
Thus $T_c'$ can be viewed as the critical temperature of the O(2)
SDW (the Kosterlitz-Thouless transition critical temperature),
which depends on the external Zeeman field $h$ as \beqn T_c' \sim
\frac{\rho_s}{ \log(\rho_s/h) }, \label{tcm}\eeqn which is
consistent with previous studies with magnetic systems with weak
anisotropy~\cite{XY}.

\begin{figure}[tbp]
\begin{center}
\includegraphics[width=180pt]{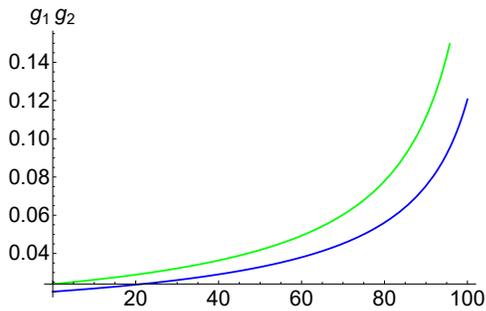}
\caption{The RG flow of $g_1$ and $g_2$ described by Eq.~\ref{RG},
with initial values $g_1 = 0.24$, $g_2 = 0.02$, and their
difference is amplified under RG. The horizonal axis is the RG
scale $l$.} \label{RGflow}
\end{center}
\end{figure}

As an illustration of the physics discussed above, let us consider
a simple case without reflection symmetry (the reflection symmetry
takes $z_\alpha \rightarrow \epsilon_{\alpha\beta}z^\ast_\beta$ in
Ref.~\cite{senthilchubukov}), where an external field leads to the
following anisotropic NLSM: \beqn \int d^2x \ \frac{1}{2g_1}
|\vect{\nabla} z_1|^2 + \frac{1}{2g_2} |\vect{\nabla} z_2|^2 +
\frac{m^2}{k_B T} |z_1|^2, \label{nlsm}\eeqn $g_i = k_B T/\rho_i$.
We take $\rho_2 > \rho_1$ for $m^2 > 0$, $i.e.$ the anisotropy
favors the condensate of $z_2$, but penalizes condensate of $z_1$.

Starting with $m = 0$, the RG flow of $g_i$ is described by the
Ricci flow~\cite{nlsm1,nlsm2}: \beqn \beta(g_{ab}) = -
\frac{1}{2\pi} \mathcal{R}_{ab}, \eeqn where $g_{ab}$ is the
metric tensor of the target manifold of the NLSM, and
$\mathcal{R}_{ab}$ is the Ricci tensor (see appendix for more
details). Expanded at the ordered state with $z_1 = 0$, $z_2 \neq
0$, the Ricci flow of the metric tensor translates into the RG
flow of $g_1$ and $g_2$ in the field theory Eq.~\ref{nlsm} \beqn
\frac{d g_1}{d\ln l} &=& \frac{1}{2\pi}\left( g_1^2 +
\frac{g_1^3}{g_2} \right) + O(g_i^3), \cr\cr \frac{d g_2}{d\ln l}
&=& \frac{1}{\pi}g_1 g_2 + O(g_i^3). \label{RG} \eeqn The RG
equations Ref.~\ref{RG} can be solved exactly for arbitrary
initial values of $g_1$, $g_2$ with a complicated form (see
appendix). If we start with a choice of different $g_1$ and $g_2$,
their difference will be amplified under RG flow
(Fig.~\ref{RGflow}). Intuitively, as long as $g_i$ are small
enough (for low enough temperature or the spin stiffness is
sufficiently strong), once $m^2$ is renormalized strong, $z_1$
will be explicitly gapped, and $z_2$ becomes an O(2) order
parameter, and enters a quasi long range algebraic phase where
$g_2$ stops growing under RG. The general physics discussed in
this section becomes manifest in model Eq.~\ref{nlsm}.

With increasing magnetic field $h$, the Mott insulator phase will
eventually be destroyed, so will the SDW order. With strong field,
the order-disorder transition of the SDW is likely a quantum phase
transition with dynamical exponent $2$, due to the precession of
the inplane SDW order parameter under an external field.

{\it --- Connections to more experimental phenomena, and
comparison with other theories}

In our picture the weak Mott insulator is a consequence of a SDW
at finite momentum, which significantly reduces the density of
``active" fermions around the Fermi surface with lowering
temperature. Thus the SDW is a competing order of the SC. We
expect this to be still true under small doping away from the Mott
insulator. Experimentally the Hall density of charge carriers in
the hole-doped Mott insulator is indeed proportional to the dopant
density away from the Mott insulator, suggesting the persistence
of the SDW under hole doping. And with an external field, either
inplane or out-of-plane, the SDW will be stabilized (the effect of
a weak magnetic field will be strongly amplified due to the
logarithmic dependence of $h$ in Eq.~\ref{tcm}), thus the SC (even
a spin triplet SC) will be significantly weakened due to its
competition with the magnetic order.

We would also like to point out that the main phenomena (1) and
(2) mentioned in the introduction are less likely to be
simultaneously consistent with other theories proposed so far.
Ref.~\onlinecite{kivelson} proposed a nematic order which
spontaneously breaks the symmetry of the valley space in the Mott
insulator phase, while Ref.~\onlinecite{lee} proposed a valence
bond solid (VBS) order in the Mott insulator. The valley space
does not have a SU(2) symmetry, hence at low temperature it would
form either a true long range order (which spontaneously breaks a
discrete symmetry) or a quasi long range order (which
spontaneously breaks the approximate U(1) valley symmetry). In
either case, it seems difficult to reconcile phenomena (1) and
(2): since the system is clearly an insulator when the SC is
suppressed, there must be no Fermi pockets left with the valley
order; but the correlation length of the valley order remains
infinite after the field is removed due to the lower symmetry of
the valley space, hence the density of ``active fermions" is still
zero, and there seems no natural way to explain the emergence of
SC inside the MI. The VBS order proposed in Ref.~\onlinecite{lee}
has the similar issue.

{\it --- Summary}

In summary we have proposed a phenomenological understanding of
the unusual emergent superconductivity inside a weak Mott
insulator observed recently in the bilayer Graphene Moir\'e
superlattice. In our picture this peculiar phenomenon is due to
the pure two dimensional nature of the system, and also the
symmetry of the order parameter that leads to the MI. We expect
this to be a quite generic mechanism, and similar behaviors can be
found in other two dimensional systems.

\begin{acknowledgments}

CX is supported by the David and Lucile Packard Foundation. The
authors thank Leon Balents, Charles Kane for very helpful
discussions. While completing this paper, we became aware of an
independent work~\cite{fu2} which aims to understand the same
experimental phenomena.

\end{acknowledgments}

\appendix

\bibliography{Graphene}

\begin{center}
\textbf{\large Appendix: From Ricci flow to RG equation}
\end{center}

In this appendix, we discuss the effect of anisotropy on the
noncollinear spin density wave from a geometric point of view. As
we argued in the main text, the ground state manifold of the
noncollinear spin density wave is a three dimensional manifold,
which will be deformed by the Zeeman field. Thus the noncollinear
spin density wave can be generally described by the NLSM \beqn
\mathcal{S}\left[X\right]=\int \frac{1}{2}
G_{ab}\left[X\right]dX^{a}\land\star dX^{b}+\ldots \label{nlsmX}
\eeqn where the bosonic field $X$ is introduced as \beqn
\left(\begin{array}{c}
X^{1}\\
X^{2}\\
X^{3}\\
\sqrt{1-\left|X\right|^{2}}
\end{array}\right)=\left(\begin{array}{c}
\textrm{Re}z_{1}\\
\textrm{Im}z_{1}\\
\textrm{Re}z_{2}\\
\textrm{Im}z_{2}
\end{array}\right),
\eeqn and the metric $G_{ab}$ should carry the information of the
external Zeeman field which lowers the symmetry of the system. In
our choice here, $X_i =0$ corresponds to the ground state
$\left|z_{1}\right|^{2}=0,\left|z_{2}\right|^{2}=1$.

To describe the geometric evolution of the target manifold more
precisely, we need to introduce our conventions of geometric
quantities. The affine connection is defined as \beqn
\Gamma_{\;bc}^{a} = \frac{1}{2}G^{ae}
\left(-\partial_{e}G_{bc}+\partial_{c}G_{be}+\partial_{b}G_{ce}\right),
\label{affine connection} \eeqn where
$\partial_{a}=\frac{\partial}{\partial X^{a}}$ is the derivative
with respect to the field $X_i$. This connection gives the Riemann
curvature \beqn R_{\;bcd}^{a}=
\partial_{c}\Gamma_{\;db}^{a}-
\partial_{d}\Gamma_{\;cb}^{a} +\Gamma_{\;ce}^{a}\Gamma_{\;db}^{e}-\Gamma_{\;de}^{a}\Gamma_{\;cb}^{e},
\eeqn and its contraction \beqn \mathcal{R}_{ab}=R_{\;acb}^{c}
\eeqn is called the Ricci tensor. The action Eq.~\ref{nlsmX} is
invariant under coordinate transformations which preserve the
distance $G_{ab}dX^{a}dX^{b}$.

Friedan~\cite{nlsm1,nlsm2} proved that the one-loop beta function
of $G_{ab}$ corresponds to the Ricci flow \beqn \frac{d
G_{ab}}{d\ln l}=-\frac{1}{2\pi}\mathcal{R}_{ab}+\ldots
\label{Ricci flow} \eeqn Then the central task is to explore how
the external Zeeman field affects the Ricci flow. Let us first
consider the simpler case without the Zeeman field. The metric
$G_{ab}$ obtained from the isotropic $\textrm{O}\left(4\right)$
NLSM reads \beqn
G_{ab}\left[X\right]=\frac{1}{g}\left(\delta_{ab}+\frac{X^{a}X^{b}}{1-\left|X\right|^{2}}\right),
\eeqn The Ricci tensor is given by \beqn
\mathcal{R}_{ab}\left[X\right]=2gG_{ab}\left[X\right], \eeqn which
is proportional to the metric. Using Eq.~\ref{Ricci flow}, we
obtain the RG flow Eq.~\ref{RG0} of the coupling constant $g$.

After turning on the Zeeman term, the $\textrm{O}\left(4\right)$
symmetry is broken, and the NLSM is modified as Eq.~\ref{nlsm}.
The metric now becomes \beqn
G_{ab}\left[X\right]=\left(\begin{array}{ccc}
\frac{1}{g_{1}} & 0 & 0\\
0 & \frac{1}{g_{1}} & 0\\
0 & 0 & \frac{1}{g_{2}}
\end{array}\right)+\frac{1}{g_{2}}\frac{X^{a}X^{b}}{1-\left|X\right|^{2}}.
\eeqn The complete expression of the Ricci tensor in this case is
rather complicated. To read the RG flow of $g_{1}, g_{2}$ from the
Ricci flow, we consider the Ricci tensor at point $X_i=0$, which
corresponds to the ordering of $z_2$ at zero temperature, and it
is the order favored by the Zeeman field: \beqn
\mathcal{R}_{ab}\left[X\rightarrow0\right]=\left(\begin{array}{ccc}
1+\frac{g_{1}}{g_{2}} & 0 & 0\\
0 & 1+\frac{g_{1}}{g_{2}} & 0\\
0 & 0 & \frac{2g_{1}}{g_{2}}
\end{array}\right).
\eeqn Combining with the value of  $G_{ab}\left[X\right]$ at $X_i
= 0$, we obtain the RG flow Eq.~\ref{RG} of $g_1$ and $g_2$.

If we start with initial values $g_1 = g$ and $g_2 = (1 - \alpha)
g$, the solution of the RG equation Eq.~\ref{RG} reads \beqn
g_1(l) &=& \frac{\pi g}{\pi - g \ln l} \cr\cr &+& \frac{g
\pi^{3/2} \left(- \pi + g \ln l + \sqrt{\pi(\pi - g \ln l)}
\right) \alpha}{\left(\pi - g \ln l \right)^{5/2}} + O(\alpha^2),
\cr\cr\cr g_2(l) &=& \frac{\pi g}{\pi - g \ln l} \cr\cr &+&
\frac{g \left(\pi^2 - 2\pi^{3/2} \sqrt{\pi - g \ln l}
\right)\alpha}{\left(\pi - g \ln l \right)^2} + O(\alpha^2). \eeqn

%Comparing the two cases, we can see the Zeeman field changes the
%geometric evolution qualitatively. Without the Zeeman field, the
%target manifold will keep being isotropic under Ricci flow.
%However, with the Zeeman field, the target manifold will become
%more and more anisotropic. This behavior strongly suggests that
%the three dimensional target manifold tends to be compactified
%into $S^1$. The final fate of the target manifold is also
%confirmed from the other fact that $z_1$ will be fully gapped when
%when $h$ is renormalized strong.

\bibliography{Graphene}

\end{document}